%% file: main.tex
\documentclass[runningheads]{llncs}
\usepackage[T1]{fontenc}

\usepackage{graphicx}
\usepackage{multirow}
\usepackage{booktabs}       
\usepackage{amsfonts}       
\usepackage{nicefrac}       
\usepackage{microtype}      
\usepackage{xcolor}         
\usepackage{graphicx}
\usepackage{colortbl}
\usepackage{float}

\usepackage[export]{adjustbox}
\usepackage{wrapfig}
\usepackage{svg}
\usepackage{amssymb}
\usepackage{sidecap}
\usepackage{caption}
\usepackage{soul}
\newcommand{\RNum}[1]{\uppercase\expandafter{\romannumeral #1\relax}}
\usepackage{placeins}  
\usepackage{amsmath}
\usepackage{caption}
\usepackage{appendix}

\usepackage{pifont}
\newcommand{\cmark}{\ding{51}}%
\newcommand{\xmark}{\ding{55}}%

\usepackage{enumitem}
\usepackage{amsfonts}       
\usepackage[colorlinks=true,linkcolor=black,citecolor=black,urlcolor=blue,bookmarks=false,hypertexnames=true]{hyperref} 
\usepackage{color}
\usepackage{tabularray}

\begin{document}
\title{Fill the K-Space and Refine the Image:\\ Prompting for Dynamic and Multi-Contrast MRI Reconstruction}
%
\titlerunning{Fill the K-Space and Refine the Image}
%
%
\author{Bingyu Xin\inst{1} \and
Meng Ye\inst{1} \and
Leon Axel\inst{2} \and
Dimitris N. Metaxas\inst{1} 
}
\authorrunning{B. Xin et al.}
\institute{Department of Computer Science, Rutgers University, Piscataway, NJ 08854, USA \and Department of Radiology, New York University, New York, NY 10016, USA\\
\email{\{bx64, my389, dnm\}@cs.rutgers.edu}\\
\url{https://github.com/hellopipu/PromptMR}}

\maketitle              

\input{content/abstract.tex}

\input{content/introduction.tex}

\input{content/prelimilaries.tex}

\input{content/method.tex}

\input{content/experiments.tex}

\input{content/conclusion.tex}

\FloatBarrier
\bibliographystyle{splncs04}
\bibliography{ref}

\input{content/appendix.tex}




\end{document}

%% file: content/abstract.tex
\begin{abstract}
The key to dynamic or multi-contrast magnetic resonance imaging (MRI) reconstruction lies in exploring inter-frame or inter-contrast information. 
Currently, the unrolled model, an approach combining iterative MRI reconstruction steps with learnable neural network layers, stands as the best-performing method for MRI reconstruction. However, there are two main limitations to overcome: firstly, the unrolled model structure and GPU memory constraints restrict the capacity of each denoising block in the network, impeding the effective extraction of detailed features for reconstruction; 
secondly, the existing model lacks the flexibility to adapt to variations in the input, such as different contrasts, resolutions or views, necessitating the training of separate models for each input type, which is inefficient and may lead to insufficient reconstruction.
In this paper, we propose a two-stage MRI reconstruction pipeline to address these limitations. The first stage involves \emph{filling the missing k-space data}, which we approach as a physics-based reconstruction problem. 
We first propose a simple yet efficient baseline model, which utilizes adjacent frames/contrasts and channel attention to capture the inherent inter-frame/-contrast correlation.
Then, we extend the baseline model to a prompt-based learning approach, \textbf{PromptMR}, for all-in-one MRI reconstruction from different views, contrasts, adjacent types, and acceleration factors.
The second stage is to \emph{refine the reconstruction} from the first stage, which we treat as a general video restoration problem to further fuse features from neighboring frames/contrasts in the image domain.
Extensive experiments show that our proposed method significantly outperforms previous state-of-the-art accelerated MRI reconstruction methods.

\keywords{ MRI reconstruction \and Prompt-based learning \and Dynamic \and Multi-contrast \and Two-stage approach}
\end{abstract}

%% file: content/introduction.tex
\section{Introduction}
Cardiovascular disease, including conditions such as coronary artery disease, heart failure, and arrhythmias, remains the leading cause of death globally. Cardiac magnetic resonance (CMR) imaging is the most accurate and reliable non-invasive technique for accessing cardiac anatomy, function, and pathology~\cite{rajiah2023cardiac}. 
In the field of accelerated MR imaging (MRI) reconstruction, unrolled networks have achieved state-of-the-art performance.
This is attributed to their ability to incorporate the known imaging degradation processes, the undersampling operation in k-space, 
into the network and to learn image priors from large-scale data~\cite{sriram2020end,fabian2022humus}. As transformers have become predominant in general image restoration tasks \cite{zamir2022restormer,liang2021swinir}, there is a noticeable trend towards incorporating transformer-based denoising blocks into the unrolled network~\cite{fabian2022humus}, which enhances reconstruction quality. However, the adoption of transformer blocks concurrently increases the network parameters and computational complexity. The stacking of denoising blocks, in an unrolled manner, further exacerbates this complexity, making the network training challenging.
Therefore, one challenging question is how to design efficient denoising blocks within an unrolled model while fully leveraging the k-space information. Another challenge arises from the versatility of MRI, which enables the acquisition of multi-view, multi-contrast, multi-slice, and dynamic image sequences, given specific clinical demands. 
While there is a prevailing trend towards designing all-in-one models for natural image restoration~\cite{li2022all,potlapalli2023promptir}, existing MRI reconstruction models cannot offer a unified solution for diverse input types. We thus endeavor to address these challenges with the following contributions:
\begin{itemize}[label=$\bullet$]
    \item Firstly, we propose a simple yet efficient convolution-only baseline model for MRI reconstruction, which outperforms previous state-of-the-art methods on two public multi-coil MRI reconstruction tasks, the CMRxRecon and fastMRI knee image reconstruction.
    \item Then, by extending our baseline model with prompt-based learning, we are the first to propose an all-in-one approach, \textbf{PromptMR}, for multi-view/-contrast and dynamic MRI reconstruction.
    \item Lastly, we extend our approach to address the capacity limitations of unrolled models, by proposing a two-stage MRI reconstruction pipeline. In the  first stage we solve a physics-based inverse problem in k-space domain to fill the missing k-space data, and in the second stage we solve a video restoration problem in the image domain to further refine the MRI reconstruction.
\end{itemize}

%% file: content/prelimilaries.tex
\section{Preliminaries}
Consider  reconstructing a complex-valued MR image $x$ from the multi-coil undersampled measurements $y$ in k-space, such that,
\begin{equation}
\label{eq:1}
y=Ax+\epsilon,
\end{equation}
where $A$ is the linear forward complex operator which is constructed based on multiplications with the sensitivity maps $S$, application of the 2D Fourier transform $F$, while it under-samples the k-space data with a binary mask $M$; $\epsilon$ is the acquisition noise. 
According to compressed sensing theory~\cite{donoho2006compressed}, we can estimate $x$ by formulating an optimization problem:
\begin{equation}
\label{eq:2}
\min_x{\frac{1}{2}||y-Ax||_2^2+\lambda R(x)},
\end{equation}
where $||y-Ax||_2^2$ is the data consistency term, $R(x)$ is a sparsity regularization term on $x$ (e.g., total variation) and $\lambda$ is a hyper-parameter which controls the contribution weights of the two terms.
E2E-VarNet~\cite{sriram2020end} solves the problem in Eq.~\ref{eq:2} by applying an iterative gradient descent method in the k-space domain. In the $t$-th step, the k-space is updated from $k^t$ to $k^{t+1}$ using:
\begin{equation}
\label{eq:3}
k^{t+1}=k^t-\eta^tM(k^t-y)+G(k^t),
\end{equation}
where $\eta^t$ is a learned step size and $G$ is a learned function representing the gradient of the regularization term $R$. We can unroll the iterative updating algorithm to a sequence of sub-networks, where each cascade represents an unrolled iteration in Eq.~\ref{eq:3}. 
The regularization term is applied in the image domain:
\begin{equation}
\label{eq:4}
G(k) = F(\mathcal{E}(\mathbf{D}(\mathcal{R}(F^{-1}(k))))),
\end{equation}
where $\mathcal{R}(x_1,...,x_N)=\sum_{i=1}^N \hat{S}_i^*x_i$ is the reduce operator that combines $N$ coil images $\{x_i\}_{i=1}^N$ via estimated sensitivity maps $\{\hat{S}_i\}_{i=1}^N$, $\hat{S}_i^*$ is the complex conjugate of $\hat{S}_i$, and $\mathcal{E}(x)=(\hat{S}_ix,...,\hat{S}_Nx) $ is the expand operator that computes coil images from image $x$. Therefore, the linear forward operator $A$ is computed as $A=MF\mathcal{E}$. $\mathbf{D}$ is a denoising neural network used to refine the complex image. $\hat{S}=\text{SME}(y_\text{ACS})$ is computed by a sensitivity map estimation (SME) network from the low-frequency region of k-space $y_\text{ACS}$, called the Auto-Calibration Signal (ACS), which is typically fully sampled. The final updated multi-coil k-space is converted to the image domain by applying an inverse Fourier transform followed by a root-sum-of-squares (RSS) method reduction~\cite{roemer1990nmr} for each pixel.

%% file: content/method.tex
\begin{figure*}[t!]
\centering
\includegraphics[width=1.0\textwidth]{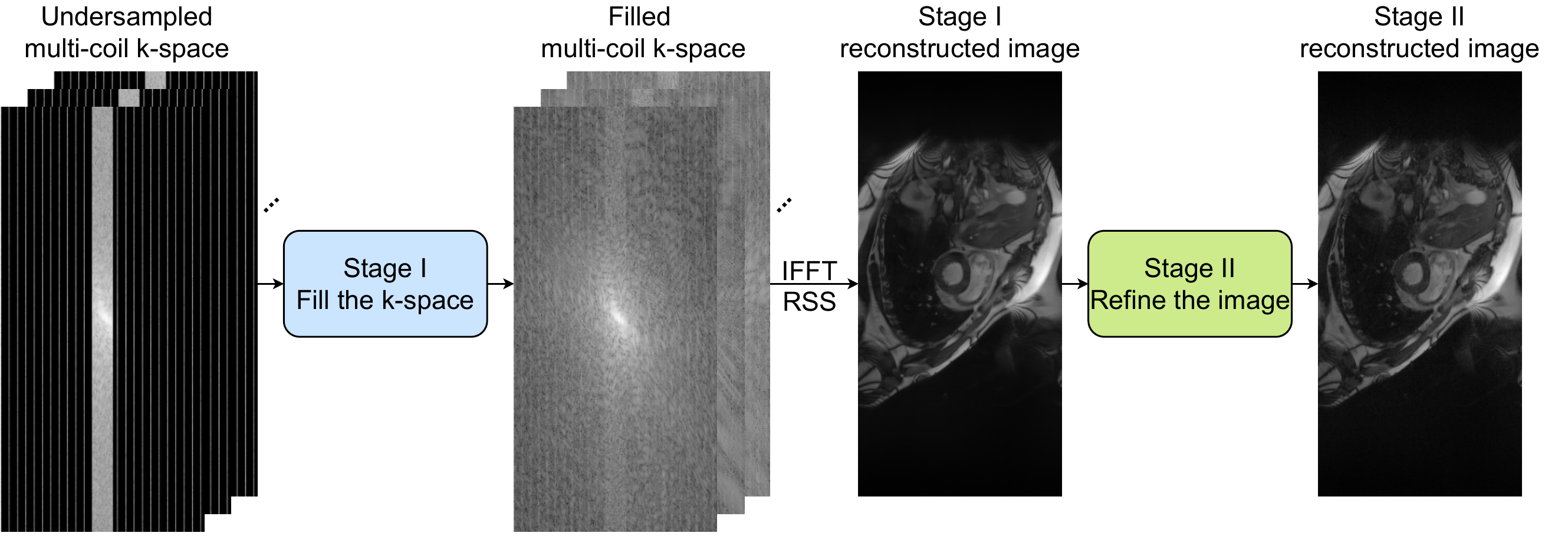}
\caption{The proposed two-stage MRI reconstruction pipeline. The first stage solves a physics-based inverse problem to fill the missing k-space data, which are then transformed to the image domain by the inverse Fast Fourier Transformation (IFFT) and root-sum-of-squares (RSS) is applied to get the first-stage reconstructed image. The second stage solves a general denoising problem to further refine the image reconstruction result.}

\label{fig:pipeline}
\end{figure*}

\section{Method}

\begin{figure*}[t!]
\centering
\includegraphics[width=1.0\textwidth]{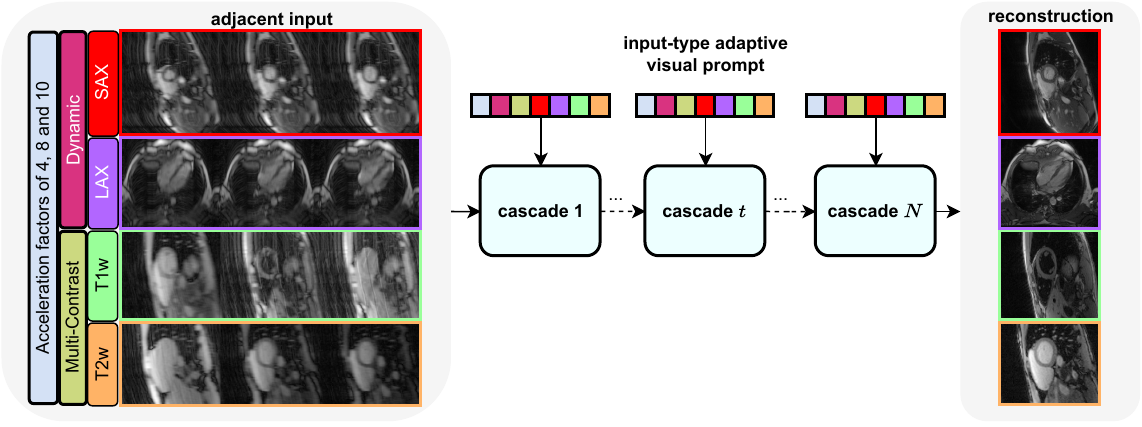}
\caption{Overview of PromptMR in Stage \RNum{1}: an all-in-one unrolled model for MRI reconstruction. Adjacent inputs, depicted in image domain for visual clarity, provide neighboring k-space information for reconstruction. To accommodate different input varieties, the input-type adaptive visual prompt is integrated into each cascade of the unrolled architecture to guide the reconstruction process.}
\label{fig:all-in-one}
\end{figure*}

We propose a two-stage pipeline for dynamic and multi-contrast MRI reconstruction, as shown in Fig.~\ref{fig:pipeline}. Below, we give more details of each stage.

\subsection{Stage \RNum{1}: Filling the K-Space}
The center of the k-space preserves image contrast, and the periphery of the k-space contains edge information. In the first stage, we fill the missing k-space data constrained by the existing k-space acquisition and learned image priors.

\subsubsection{Baseline Model}
We follow the implementation of E2E-VarNet \cite{sriram2020end} to construct an unrolled model in Stage \RNum{1}.
Inspired by the adjacent slice reconstruction (ASR) method~\cite{fabian2022humus}, which learns inter-slice information by jointly reconstructing a set of adjacent slices instead of relying on a single k-space to be reconstructed, we devise the following new method. We generalize ASR to adjacent k-space reconstruction along any dimension, e.g., temporal/slice/view/contrast dimension, and the updating formula of Eq.~\ref{eq:3} is improved as follows:
\begin{equation}
\label{eq:5}
k^{t+1}_{adj}=k^t_{adj}-\eta^tA(k^t_{adj}-y_{adj})+G(k^t_{adj}),
\end{equation}
where $k^t_{adj}=[k^t_{c-a}, ..., k^t_{c-1}, k^t_{c}, k^t_{c+1}, ..., k^t_{c+a}]$ is the concatenation of the central k-space $k^t_{c}$ with its $2a$ adjacent k-spaces along a specific dimension. 
To efficiently extract features from adjacent inputs, we design a Unet-style network~\cite{ronneberger2015u} with channel attention~\cite{hu2018squeeze,huang2019mri}, namely CAUnet,  for both the denoising network $D$ and the sensitivity map estimation network, as shown in Appendix~\ref{app:ca-unet}. The CAUnet has a 3-level encoder-decoder structure. Each level consists of a DownBlock, UpBlock, and corresponding skip connection. The architecture integrates a BottleneckBlock for high-level semantic feature capturing and employs Channel Attention Blocks (CABs) within each block. The overall unrolled architecture is shown in Appendix~\ref{app:unrolled_model}.

\begin{figure*}[t!]
\centering
\includegraphics[width=1.0\textwidth]{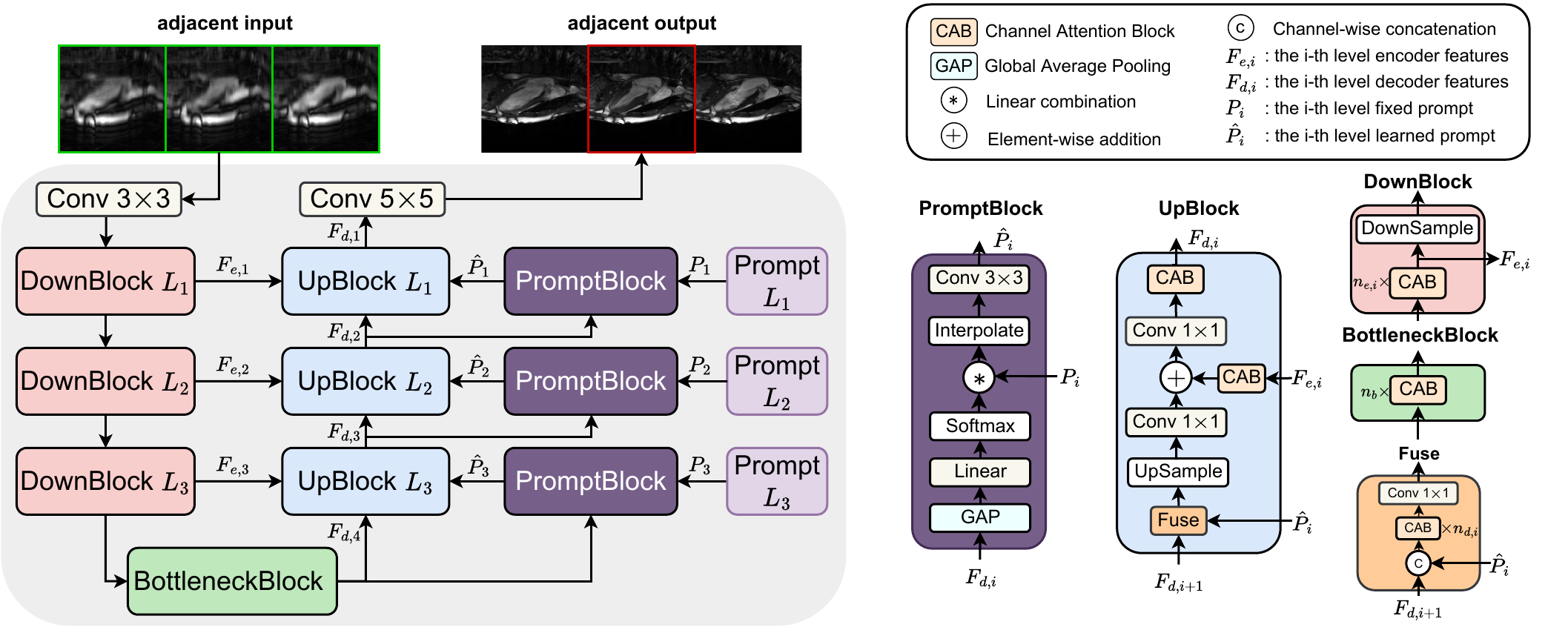}
\caption{Overview of the PromptUnet architecture in PromptMR, featuring a 3-level encoder-decoder design. Each level comprises a DownBlock, UpBlock and PromptBlock. The PromptBlock in the $i$-th level encodes input-specific context into fixed prompt $P_i$, producing adaptively learned prompt $\hat{P}_i$. These prompts, across multiple levels, integrate with decoder features $F_{d,i}$ in the UpBlocks to allow rich hierarchical context learning.}
\label{fig:promptmr}
\end{figure*}

\subsubsection{PromptMR}
Considering various image types (e.g., different views, different contrasts) with different adjacent types (e.g., dynamic, multi-contrast) under different undersampling rates (e.g., $\times 4, \times 8, \times 10$), instead of training separate models for each specific input, we propose to learn an all-in-one unified model for all possible adjacent inputs. 
The image structure remains consistent for multi-contrast adjacent input, while only the contrast varies. Conversely, the contrast remains constant for dynamic adjacent input, but the image structure shifts.
To achieve effective performance on diverse input types, the unified model should be able to encode the contextual information conditioned on the input type. Inspired by the recent development of visual prompt learning~\cite{jia2022visual,khattak2023maple} and prompt learning-based image restoration method~\cite{potlapalli2023promptir}, we introduce PromptMR, an all-in-one approach for MRI reconstruction, as illustrated in Fig.~\ref{fig:all-in-one}. While PromptMR retains the same unrolled architecture of the basline model, it extends CAUnet to PromptUnet by integrating PromptBlocks to learn input-type adaptive prompts and then interact with decoder features in the UpBlocks at multiple levels, to enrich the input-specific context,  as shown in Fig.~\ref{fig:promptmr}. 
The PromptBlock at $i$-th level takes features $F_{d,i}\in\mathbb{R}^{H_f\times W_f\times C_f}$ from the decoder and $N_p$-components fixed prompt $P_i\in\mathbb{R}^{N_p\times H_p\times W_p\times C_p}$ as input. Then, $F_{d,i}$ are processed by a global average pooling (GAP) layer, followed by a linear layer and a softmax layer to generate the normalized prompt weights $\{\omega_{ij}\}_{j=1}^{N_p}$. These weights linearly combine the $N_p$ prompt components as $\sum_{j=1}^{N_p} {\omega_{ij} P_{ij}}$, which is then interpolated to match the spatial dimension of $F_{d,i}$, before going through a $3\times3$ convolution layer to generate the input-type adaptive prompt $\hat{P}_i$. The process in the PromptBlock can be summarized as:
\begin{equation}
\label{eq:6}
\begin{array}{cc}
\hat{P}_{i}=\text{Conv}_{3\times 3}(\text{Interp}(\sum_{j=1}^{N_p} {\omega_{ij} P_{ij}})), &\quad  \omega_i = \text{Softmax}(\text{Linear}(\text{GAP}(F_{d,i})))
\end{array}
\end{equation}
The generated prompts by the PromptBlocks at multiple levels can learn hierarchical input-type contextual representations, which are integrated with the decoder features to guide the all-in-one MRI reconstruction.

\subsection{Stage \RNum{2}: Refining the Image}
After the first stage, the missing k-space data has been filled, and image aliasing artifacts have been largely removed. However, due to the unrolled nature and memory limitations, the capability of the denoising block we can use is constrained, which may prevent the full exploration of dynamic and multi-contrast information. In stage \RNum{2}, we further explore the inter-frame/-contrast coherence in the image domain for multi-frame/-contrast feature aggregation by using a powerful restoration model, ShiftNet \cite{li2023simple}, as the refinement network. This network employs stacked Unets and grouped spatio-temporal shift operations to expand the effective receptive fields. Details of the ShiftNet are not covered here, since it is not the core part of this paper, and ShiftNet can be replaced by any state-of-the-art video restoration model.

%% file: content/experiments.tex
\section{Experiments}
In this section, we first provide experimental details and results of our proposed method on the CMRxRecon dataset.
We use SSIM, PSNR, and NMSE to compare the performance of different reconstruction methods under various acceleration factors ($\times 4$, $\times 8$, $\times 10$). Then, we conduct extensive ablation studies of our proposed method and also benchmark on another large-scale MRI dataset, the fastMRI multi-coil knee dataset. For experiments on fastMRI dataset, we refer readers to the Appendix~\ref{app:add_exp}.

\subsection{CMRxRecon Dataset}
The CMRxRecon Dataset~\cite{wang2023cmrxrecon} includes 120 cardiac MRI cases of fully sampled dynamic cine and multi-contrast raw k-space data obtained on 3 Tesla magnets. The dynamic cine images in each case include short-axis (SAX), two-chamber (2-CH), three-chamber (3-CH), and four-chamber (4-CH) long-axis (LAX) views. Typically $5\sim10$ slices were acquired for SAX cine, while a single slice was acquired for each LAX view. The cardiac cycle was segmented into $12\sim25$ phases with a temporal resolution of 50 ms. The multi-contrast cardiac MRI in each case is in the SAX view, which contains 9 T1-weighted (T1w) images conducted using a modified look-locker inversion recovery (MOLLI) sequence and 3 T2-weighted (T2w) images performed using T2-prepared FLASH sequence. 

The shape of each k-space data is [time phases/contrasts, slices, coils, readouts, phase encodings]. 
All data were compressed into 10 virtual coils. We splited the cases in an $8\, :\, 2$ ratio, resulting in 14, 964 dynamic images and 6, 516 multi-contrast images for training, and 2, 940 dynamic images and 1, 272 multi-contrast images for testing.

\subsection{Results}
We assessed the performance of our proposed baseline model, PromptMR, and two-stage reconstruction pipeline using the CMRxRecon dataset. In the first stage, we compared the E2E-VarNet~\cite{sriram2020end} and HUMUS-Net-L~\cite{fabian2022humus} with our baseline in a one-by-one setup, in which we trained four separate models from scratch for SAX/LAX/T1w/T2w reconstruction task, respectively. Then we compared our PromptMR and PromptIR~\cite{potlapalli2023promptir} in an all-in-one configuration. 
In the second stage, we deployed ShiftNet to refine the images reconstructed by PromptMR. In our experiment,  we minimize the SSIM loss between the target image and the reconstructed image; all unrolled models consist of 12 cascades, except for HUMUS-Net-L, which only has 8 cascades due to its large parameter size; we trained networks using AdamW~\cite{loshchilov2017decoupled} optimizer with a weight decay of 0.01 for 12 epochs; the learning rate was set as $2\times 10^{-4}$ for the first 11 epochs and $2\times 10^{-5}$ for the last epoch. 

The results are shown in Table~\ref{tab:benchmark}. Notably, our baseline model outperforms E2E-VarNet and HUMUS-Net-L across all tasks. Moreover, our PromptMR demonstrates significant enhancement in the all-in-one setup when compared to the baseline model trained for individual tasks. PromptIR performs poorly due to the fact that it is not tailored to account for the MRI forward model. The refinement in the second stage offers a marginal boost to the SSIM, but provides considerable improvements for NMSE and PSNR. The qualitative results are shown in Fig.~\ref{fig:fig_lax}. More qualitative comparisons can be found in Appendix~\ref{app:cmrxrecon_qualitative}. 
These qualitative comparisons show that our method can recover more finer details for small anatomical structures on the reconstructed images.

\begin{table}[t!]
  \centering
  \caption{Comparison of \ul{NMSE($\times 10^{-2}$)/PSNR/SSIM} of different MRI reconstruction methods on CMRxRecon dataset under $\times 10$ acceleration. The best and second best results are highlighted in \textcolor{red}{red} and \textcolor{blue}{blue} colors, respectively.}
  \label{tab:benchmark}
  \resizebox{\textwidth}{!}{
  \begin{tabular}{@{}c|c|c|cc|cc@{}}
    \toprule
    \multirow{2}{*}{Stage} & \multirow{2}{*}{Task} & \multirow{2}{*}{Method} & \multicolumn{2}{c|}{Cine} & \multicolumn{2}{c}{Mapping} \\
    & & & SAX & LAX  & T1w & T2w  \\
    \midrule
    \multirow{5}{*}{\RNum{1}}& \multirow{3}{*}{One-by-One} & E2E-Varnet~\cite{sriram2020end} & 1.6/42.05/0.9744 & 2.1/39.93/0.9673 & 1.5/43.12/0.9800 & 1.4/41.20/0.9777  \\
    && HUMUS-Net-L~\cite{fabian2022humus} & 1.3/42.96/0.9791 & 2.0/40.07/0.9689 & 1.3/43.85/0.9832 & 1.1/42.39/0.9824  \\
    & & Baseline (Ours) & 1.1/43.68/0.9814 & 1.9/40.38/0.9705 & 1.2/44.14/0.9839  & 0.9/43.12/0.9845   \\
    \cmidrule{2-7}
    & \multirow{2}{*}{All-in-One}& PromptIR~\cite{potlapalli2023promptir} & 2.5/40.16/0.9659 & 2.7/38.62/0.9581 & 2.3/41.10/0.9726 & 1.4/41.10/0.9784 \\
    & & PromptMR (Ours) & \textcolor{blue}{1.1/45.58/0.9865}  & \textcolor{blue}{1.2/43.72/0.9836}  & \textcolor{blue}{1.0/46.84/0.9899} & \textcolor{blue}{0.7/46.24/0.9903} \\
    \midrule
     \RNum{2} & \multicolumn{2}{c|}{PromptMR+ShiftNet~\cite{li2023simple}} & \textcolor{red}{0.7/45.63/0.9866} & \textcolor{red}{0.9/43.76/0.9837} & \textcolor{red}{0.7/47.04/0.9903} & \textcolor{red}{0.5/46.33/0.9905} \\
    
    \bottomrule
  \end{tabular}}
\end{table}

\begin{figure*}[t!]
\centering
\includegraphics[width=1.0\textwidth]{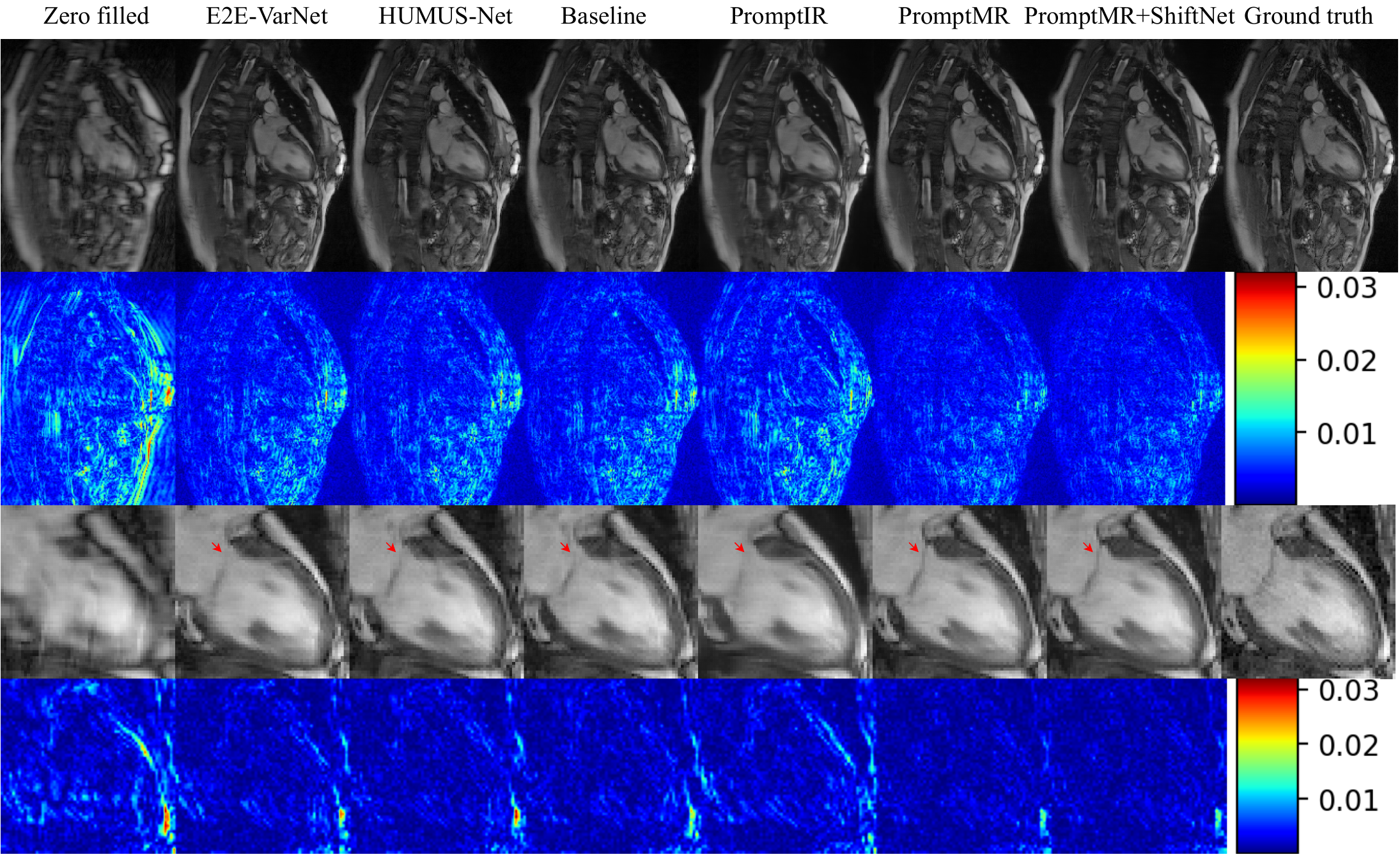}
\caption{The reconstruction results and absolute error maps of different methods for LAX 2-CH cine image of case P101 under $\times 10$ acceleration factors. The bottom two rows show the zoomed area. Red arrows show the difference in recovery of the mitral valve structure for different reconstruction methods.}
\label{fig:fig_lax}
\end{figure*}

\subsection{Ablation Study}

\subsubsection{Single MRI Reconstruction Task}
We started with an ablation study on two single MRI reconstruction tasks, dynamic cine SAX image reconstruction and multi-contrast T1-weighted image reconstruction, both under $\times 10$ acceleration, to investigate the impact of adjacent reconstruction and prompt module in the proposed PromptMR. 
We changed the number of adjacent images to 1, 3, and 5, where `1' indicates the absence of adjacent input. The results, shown in Table~\ref{tab:abalation}, underscore the utility of incorporating adjacent input to enhance the reconstruction quality. Moreover, the inclusion of PromptBlocks proves beneficial for individual MRI reconstruction tasks.

\begin{table}[t!]
    \centering
    \captionof{table}{Impact of the adjacent input number and the PromptBlock in PromptMR for two single MRI reconstruction tasks: dynamic cine SAX and multi-contrast T1-weighted (T1w) reconstruction under $\times 10$ acceleration. }
    \label{tab:abalation}
    \scalebox{1.0}{
    \begin{tabular}{c|c|c|c}
    \toprule
        \multirow{2}{*}{\# of adj}  & \multirow{2}{*}{PromptBlock} & SAX & T1w\\
        & &  PSNR/SSIM & PSNR/SSIM \\
    \midrule
        1 &  \cmark & 43.19/0.9798 & 44.36/0.9845\\
        3 &  \cmark & 43.96/0.9822 & 44.78/0.9856\\
        5 &  \cmark & 43.87/0.9820 & 44.75/0.9856\\
        5 &  \xmark & 43.68/0.9814 & 44.14/0.9839\\
    \bottomrule
    \end{tabular}}
\end{table}

\subsubsection{All-In-One MRI Reconstruction Task}
\begin{figure*}[t!]
\centering
\includegraphics[width=1.0\textwidth]{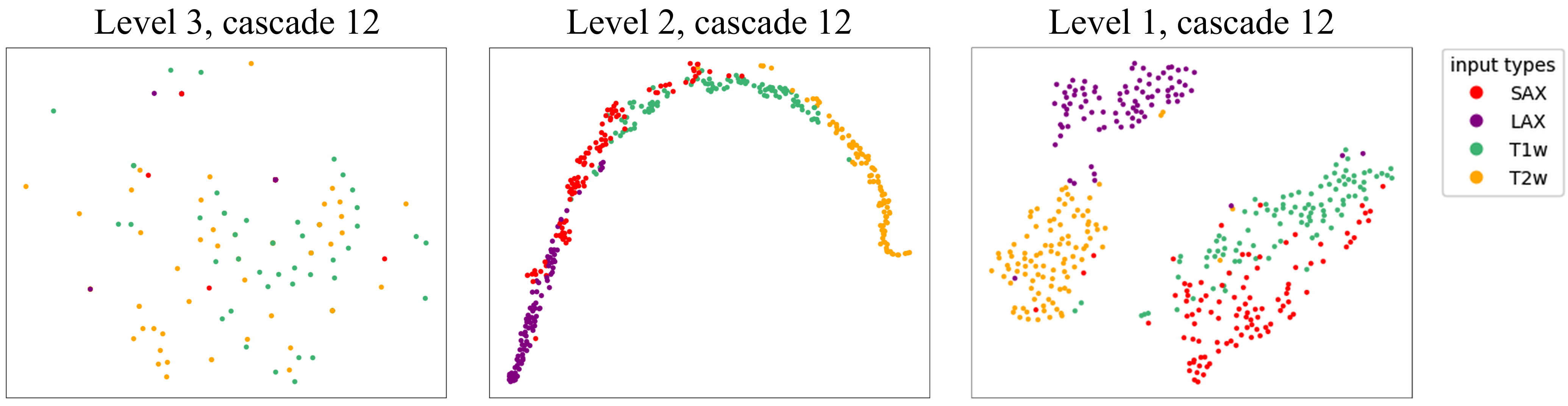}
\caption{Visualization of the learned prompts at each decoder level in the $12$-th cascade in PromptMR using t-SNE.}
\label{fig:tsne}
\end{figure*}

To investigate the impact of the PromptBlock in the all-in-one MRI reconstruction task, we trained both our baseline and PromptMR model using all possible input data in the CMRxRecon dataset. As depicted in Table~\ref{tab:abalation_all}, the integration of the PromptBlock into our baseline model enables PromptMR to achieve significant improvements across all individual reconstruction tasks. We also used t-SNE~\cite{van2008visualizing} to visualize the learned prompts in the $12$-th cascade at multiple decoder levels from different types of data in the test set. Fig.~\ref{fig:tsne} shows that the prompts can learn to encode discriminative information for different input types at lower levels.

\begin{table}[t!]
    \centering
    \captionof{table}{Impact of PromptBlock in all-in-one task. Results are reported on the CMRxRecon dataset under $\times 10$ acceleration. }
    \label{tab:abalation_all}
    \scalebox{1.0}{
    \begin{tabular}{c|c|c|c|c}
    \toprule
        \multirow{2}{*}{Method} & SAX & LAX & T1w & T2w\\
                                & PSNR/SSIM & PSNR/SSIM & PSNR/SSIM & PSNR/SSIM \\
    \midrule
        Baseline (Ours) & 43.97/0.9825  & 42.11/0.9786 & 44.90/0.9862 & 44.45/0.9874  \\
        PromptMR (Ours) & 45.58/0.9865 & 43.72/0.9836 & 46.84/0.9899 & 46.24/0.9903 \\
    \bottomrule
    \end{tabular}}
\end{table}

%% file: content/conclusion.tex
\section{Conclusion}

In this work, we introduce a robust baseline model for MRI reconstruction that utilizes neighboring information of adjacent k-space. To accommodate various input types, adjacent configurations, and undersampling rates within a unified model, we enhance our baseline with prompt-based learning blocks, creating an all-in-one MRI reconstruction model, \textbf{PromptMR}. Finally, to overcome the model capacity constraints of unrolled architectures, we propose a second stage of image refinement to delve deeper into the adjacent information, which is particularly useful when immediate reconstruction latency is not a priority.

%% file: content/appendix.tex
\newpage
\appendix
\section*{Appendix}

{

\counterwithin{figure}{section}
\counterwithin{table}{section}
\section{Details of the Baseline Model}
\label{app:baseline}

\subsection{CAUnet}
\label{app:ca-unet}

\begin{figure*}[ht]
\centering
\includegraphics[width=1.0\textwidth]{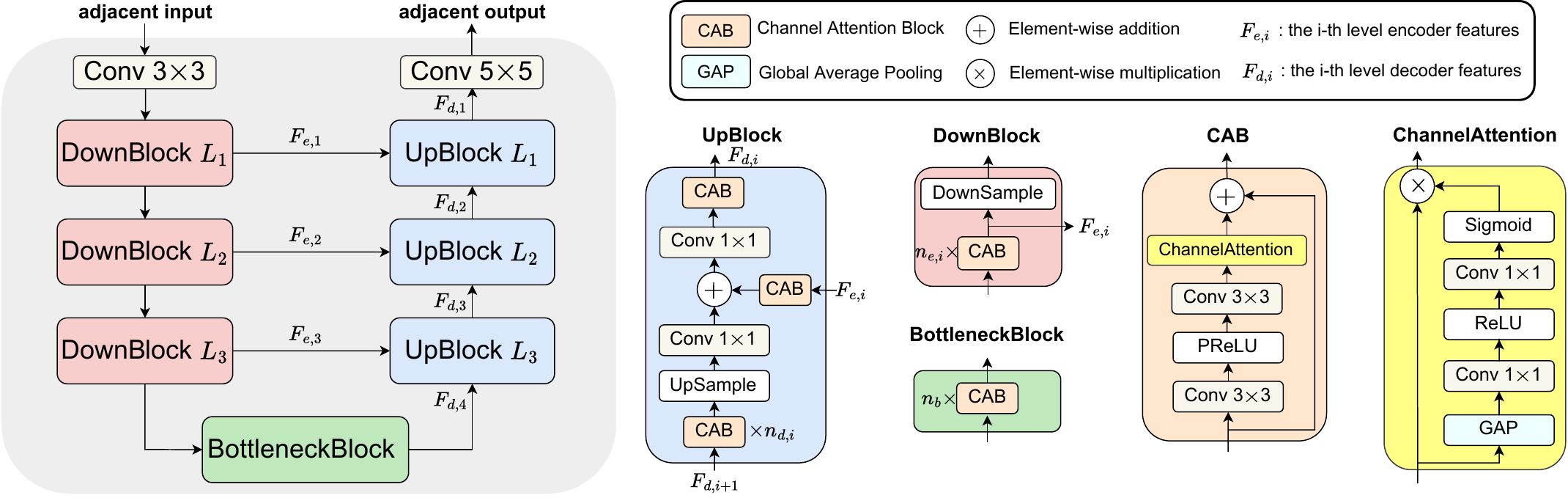}
\caption{Overview of the CAUnet architecture in the proposed baseline model.}
\label{fig:baseline}
\end{figure*}

\subsection{Unrolled model architecture}
\label{app:unrolled_model}


\begin{figure*}[ht]
\centering
\includegraphics[width=1.0\textwidth]{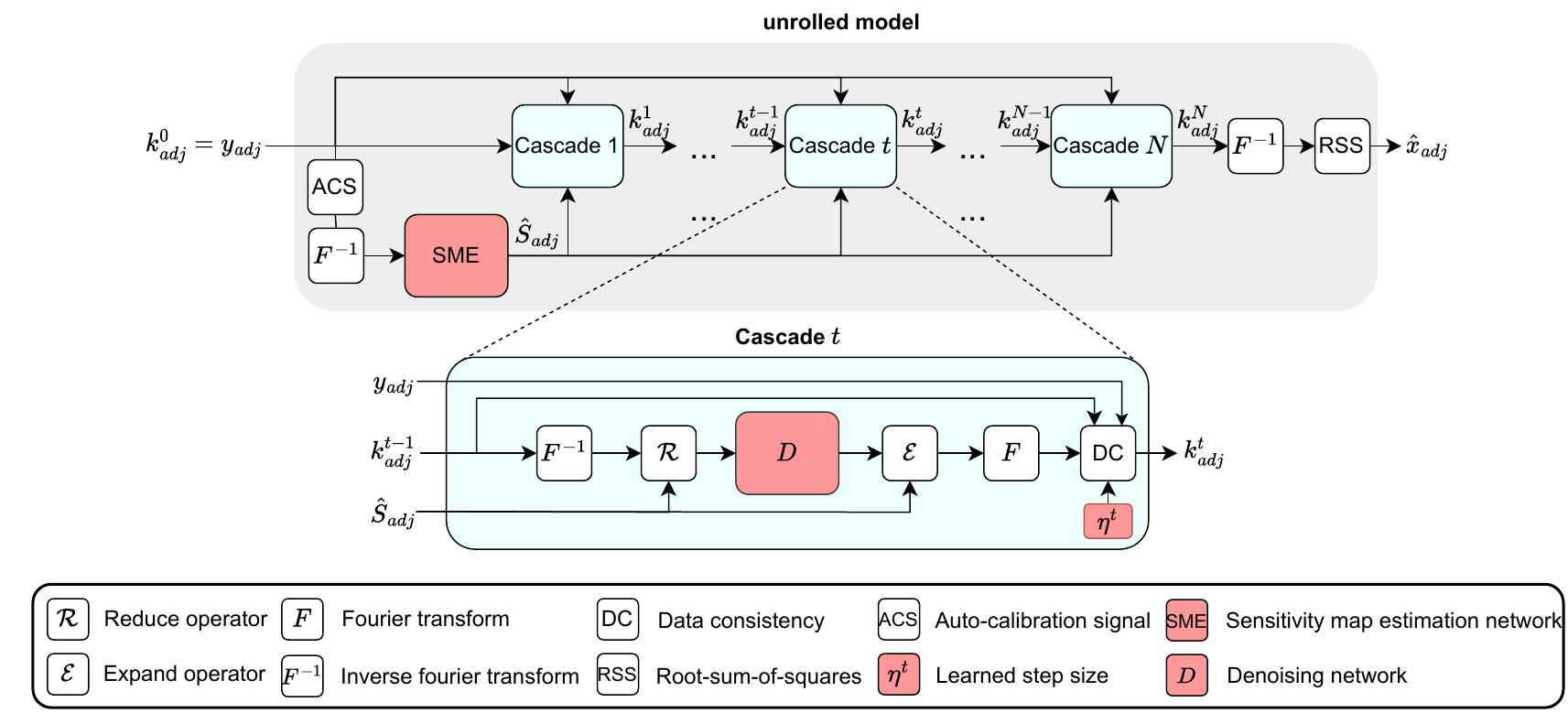}
\caption{Overview of the unrolled model architecture for both our baseline model and PromptMR. The primary distinction is in the denoiser $D$ and sensitivity map estimation (SME) networks: the baseline employs CAUnet, whereas PromptMR utilizes PromptUnet. Each cascade represents an updating step in Eq.~\ref{eq:5} in the main text. The red module indicates the learnable part in the unrolled model.}
\label{fig:unroll}
\end{figure*}

\section{Experiments on FastMRI Multi-Coil Knee Dataset}
\label{app:add_exp}
\subsubsection{Benchmark on FastMRI Multi-Coil Knee Dataset}
To assess the performance of our proposed method across different anatomies, we benchmarked it on another large-scale MRI reconstruction dataset, the fastMRI multi-coil knee dataset~\cite{zbontar2018fastmri}. Since the online evaluation platform for the fastMRI test set is unavailable\setcounter{footnote}{0}\footnote{\url{https://github.com/facebookresearch/fastMRI/discussions/293}}, we divided the original 199 validation cases into 99 for validation and 100 for testing. The results of other methods are reported using their officially pretrained models. As presented in Table~\ref{tab:fastmri}, our models outperform all previous state-of-the-art methods, without significantly increasing the number of network parameters compared to E2E-Varnet.\\

\begin{table}[h!]
    \centering
    \captionof{table}{Performance of state-of-the-art accelerated MRI reconstruction techniques on the fastMRI knee multi-coil $\times 8$ test dataset. The best and second best results are highlighted in \textcolor{red}{red} and \textcolor{blue}{blue} colors, respectively.}
    \label{tab:fastmri}
    \scalebox{1.0}{
    \begin{tabular}{c|c|c|c|c}
    \toprule
        Method & \# of params & NMSE($\times 10^{-2}$)($\downarrow$) & PSNR($\uparrow$) & SSIM($\uparrow$)\\
    \midrule
        E2E-Varnet~\cite{sriram2020end}& 30M &0.8690 $\pm$ 0.9279 & 37.30 $\pm$ 4.925 & 0.8936 $\pm$ 0.1157  \\
        HUMUS-Net~\cite{fabian2022humus}& 109M &0.8974 $\pm$ 0.9743 & 37.20 $\pm$ 5.009 & 0.8946 $\pm$ 0.1162\\
        HUMUS-Net-L~\cite{fabian2022humus}& 228M & 0.8587 $\pm$ 0.9930 & 37.45 $\pm$ 5.067 & 0.8955 $\pm$ 0.1161\\
    \midrule
        Baseline (ours)& 47.5M & \textcolor{red}{0.8321 $\pm$ 0.9258}& \textcolor{blue}{37.57 $\pm$ 5.143}&  \textcolor{blue}{0.8964 $\pm$ 0.1162}  \\
        PromptMR (ours)& 79.6M & \textcolor{blue}{0.8344 $\pm$ 0.9648}& \textcolor{red}{37.63 $\pm$ 5.319}&  \textcolor{red}{0.8970 $\pm$ 0.1168}  \\
    \bottomrule
    \end{tabular}}
\end{table}

\subsubsection{Effectiveness of Two-Stage Pipeline}
We employed ShiftNet to refine the images reconstructed by the pretrained E2E-Varnet on the fastMRI multi-coil knee test dataset with $\times 8$ undersampling. Table~\ref{tab:abalation_stage2} shows that the second-stage refinement substantially improves the reconstruction quality, which implies that the multi-slice information in the fastMRI dataset might not be comprehensively utilized by the single-stage unrolled model.\\

\begin{table}[h!]
    \centering
    \captionof{table}{Effectiveness of the second-stage image refinement on the fastMRI knee multi-coil $\times 8$ test dataset.}
    \label{tab:abalation_stage2}
    \scalebox{1.0}{
    \begin{tabular}{c|c|c|c|c|c}
    \toprule
      Stage   & Method & \# of params & NMSE($\times 10^{-2}$)($\downarrow$) & PSNR($\uparrow$) & SSIM($\uparrow$)\\
    \midrule
       \RNum{1} & E2E-Varnet~\cite{sriram2020end} & 30M &0.8690 $\pm$ 0.9279 & 37.30 $\pm$ 4.925 & 0.8936 $\pm$ 0.1157  \\
    \midrule
       \RNum{2} & ShiftNet~\cite{li2023simple} & 2M &0.8415 $\pm$ 0.9131 & 37.46 $\pm$ 4.973 & 0.8953 $\pm$ 0.1157\\
    \bottomrule
    \end{tabular}}
\end{table}

\section{Additional Qualitative Results on CMRxRecon Dataset}
\label{app:cmrxrecon_qualitative}


\begin{figure*}[h!]
\centering
\includegraphics[width=1.0\textwidth]{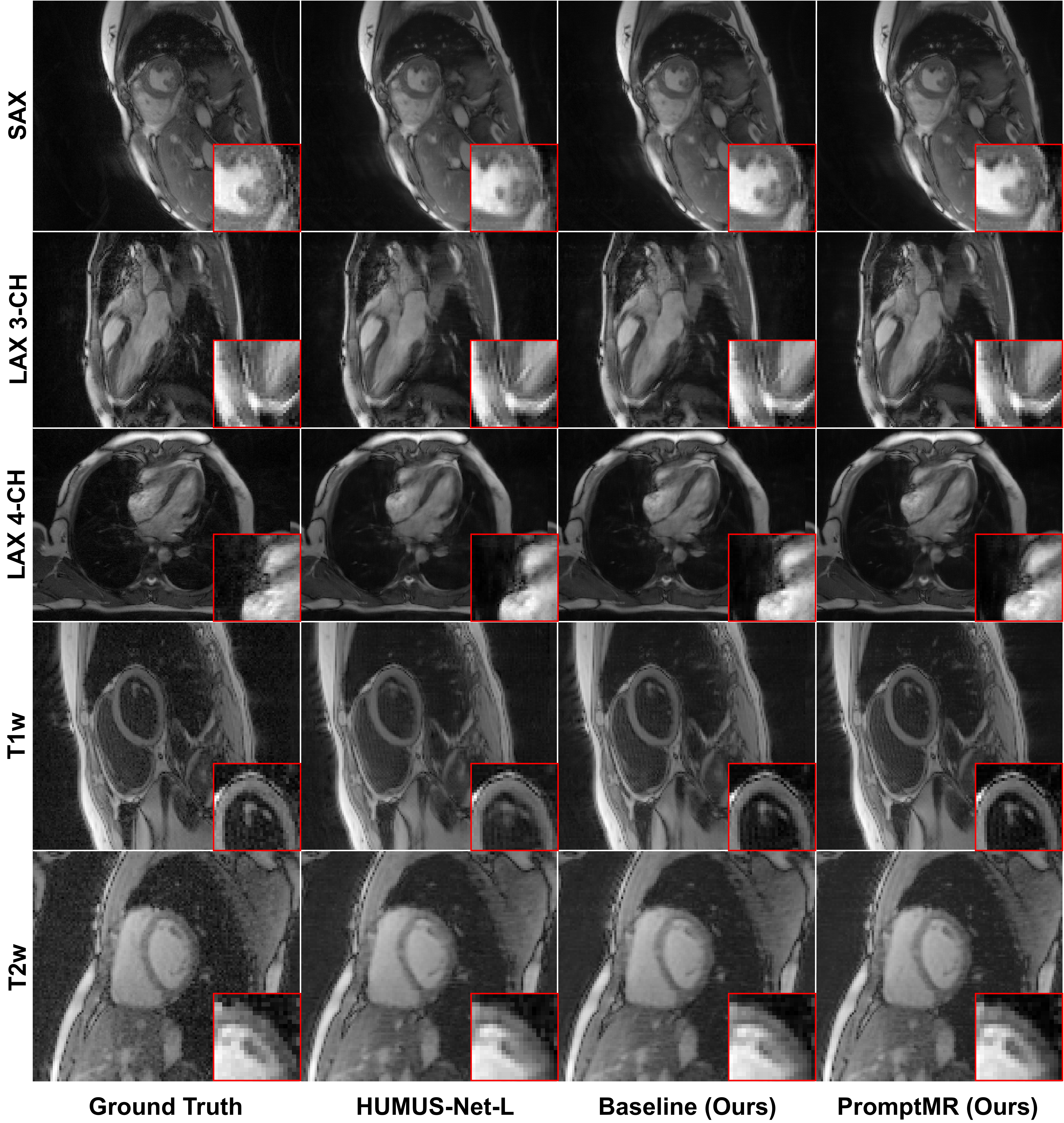}
\caption{Visual comparison of reconstructions from the CMRxRecon dataset with $\times10$ acceleration.
PromptMR can recover fine details (highlighted in red box) on reconstructed images that other state-of-the-art methods may miss.}
\label{fig:qualitative_more}
\end{figure*}

}